\documentclass{mem}
\usepackage{natbib}\usepackage{txfonts}\usepackage{balance}
\usepackage{graphicx}
\usepackage[a4paper]{hyperref}
\idline{75}{282}
\begin{document}
\def\teff{$T\rm_{eff }$}
\def\kms{$\mathrm {km s}^{-1}$}

\def\ltsim{\raise 2pt \hbox {$<$} \kern-1.1em \lower 4pt \hbox {$\sim$}}
\def\gtsim{\raise 2pt \hbox {$>$} \kern-1.1em \lower 4pt \hbox {$\sim$}}

\title{
GMRT 150 MHz follow up of diffuse steep spectrum radio emission in galaxy clusters
}

%\subtitle{}

\author{
G. \,Macario\inst{1},  
T. \, Venturi\inst{1},
D. Dallacasa\inst{1,2},
S. Giacintucci\inst{1,3},
G. Brunetti\inst{1},\\
R. Cassano\inst{1},
C. H. Ishwara-Chandra\inst{4},
R. Athreya\inst{5}	
}

  \offprints{G. Macario}

\institute{
INAF --Istituto di Radioastronomia, Via Gobetti 101,
I-40131 Bologna, Italy\\
\email{g.macario@ira.inaf.it}
\and
Dipartimento di Astronomia, Universit\'a di Bologna, 
Via Ranzani 1, I-40127 Bologna, Italy
\and
Department of Astronomy, University of Maryland, College Park, MD 20742-2421
\and
National Centre for Radio Astrophysics, TIFR, Ganeshkhind, Pune 411007, India
\and
Indian Institute of Science Education and Research, Sutarwadi Road, Pashan, 
Pune 411021, INDIA
}

\authorrunning{Macario et al.}

\titlerunning{Low frequency follow-up}

\abstract{

It has been recently found that a few galaxy clusters host diffuse 
radio halo emission with very steep synchrotron spectra ($\alpha >$ 1.6), 
which may be classified as Ultra Steep Spectrum Radio Halos (USSRHs).
USSRHs are expected in the turbulence re-acceleration model for the origin of cluster radio halos, 
and are best discovered and studied at low frequencies. 
We performed GMRT follow up observations of three galaxy clusters at 150 MHz,  
selected from the GMRT radio halo survey, which are known to host an USSRH 
or candidate very steep spectrum diffuse emission.  
This project is aimed to characterize the low frequency spectrum of USSRHs for a detailed study 
of their origin and connection with cluster mergers. 
We present preliminary results at 150 MHz of the cluster A\,697.

}

\maketitle{}

\section{Introduction}

Radio halos are diffuse Mpc-scale sources observed at the centre of 
a fraction of massive galaxy clusters. 
They have steep synchrotron spectra, 
with typical spectral index $\alpha \simeq$ 1.3-1.4 (in the convention $S\propto \nu^{-\alpha}$). 
However, recent low frequency observations of a few clusters have revealed the existence 
of radio halos with much steeper spectra 
($\alpha >$ 1.6; A\,521,\citealt{a521nature08}; A\,2256,\citealt{brentjens08}; A\,1914,\citealt{gg09}; 
A\,697,\citealt{maca10}; hereafter MVB10). \\
\indent The integrated spectrum of radio halos is a key observable to address the question of their origin. 
The turbulent re-acceleration model (e.g. \citealt{petrosian01};\citealt{brunetti01}) 
provides the unique expectation of spectra
much steeper than those found to date, as a consequence 
of the spectral steepening below GHz frequencies in the spectrum of radio halos generated in 
less energetic merger events (e.g. \citealt{cassano09}). 
The detection of radio halos with very steep spectrum ($\alpha >$ 1.6) 
would be a major piece of evidence in support of this scenario, and at the
same time it would disfavour secondary models (e.g.\citealt{blasi99}) 
for the origin of the emitting electrons 
which, for observed steep spectra, requires a very large proton energy budget (e.g.\citealt{brunetti04}). 
These giant radio halos are best studied at low frequencies, due to their very steep spectrum. 
Moreover, statistical calculations based on the re-acceleration model predicts that their 
number increases at low frequencies \citep{cassano10a}. \\
\indent The prototype of these sources, which we refer to as ultra
steep spectrum radio halos (hereinafter USSRH), was 
discovered in the merging cluster A\,521 and has a spectrum with $\alpha\simeq1.9$ 
in the frequency range 235-1400 MHz \citep{a521nature08,dallacasa09}. \\
\indent A low frequency follow-up of a few galaxy clusters selected  from the GMRT radio halo survey cluster  sample \citep{venturi07,venturi08} has been recently carried out with the GMRT at 325 and 240 MHz (\citealt{venturi09}, Giacintucci, this volume; Venturi et al. to be submitted). \\
An important result of those deep low frequency observations has been the discovery of very steep spectrum diffuse emission at the centre of three clusters. Beyond the USSRH in A\,521, another radio halo with very steep spectrum was found in the cluster A\,697 (MBV10). %\citealt{maca10}). 
Moreover, candidate very steep spectrum radio emission has been found at the centre of A\,1682 (\citealt{venturi09}, 
Venturi et al. to be submitted).\\
For these three clusters, deep GMRT follow-up observations at 150 MHz were performed, 
in order to constrain the low frequency end of the spectra of the two USSRH and to 
carry out an appropriate study of the steep spectrum diffuse emission in A\,1682. 

\section{GMRT 150 MHz observations}

In Table \ref{tab:obs} we report the main parameters of the GMRT 150 MHz observations. 
In order to achieve high sensitivity and to ensure an appropriate uv-coverage, 
each cluster was observed for a total time of 10 hours. 
All the observations were performed recording only one sideband, in the default spectral line mode, with the 8 MHz band divided into 128 channels,  
each 62.5 kHz wide. \\
\indent In this paper we present the data analysis and preliminary images for A\,697.  
Data reduction of the other observations is in progress. 
%
%%%%%%%%%%%%%%%%%%%%%%%%%%%%TAB 1%%%%%%%%%%%%%%%
\begin{table*}%[ht!]
\caption{Summary of GMRT 150 MHz observations}
\begin{center}
\begin{tabular}{lccccccc}
\hline
Cluster name & RA$_J2000$ & DEC$_J2000$ & z & Obs. Date &$\nu$ & $\Delta\nu$ & Obs. time \\
 			      & 			    & 			     &    &                  &(MHz) & (MHz)        & (hours) \\
\hline
A\,0521  	& 04 54 09.1 &  -10 14 19  & 0.2475  &2009,  Aug 16  & 151 &  8 & 10 \\
A\,1682  & 13 06 49.7 & +46 32 59 & 0.2260  & 2009, Aug 17 & 151  & 8 & 10 \\
A\,0697 	& 08 42 53.3 & +36 20 12  & 0.2820  & 2009, Aug 30 &  151 & 8 & 10 \\
\hline
\end{tabular}
\end{center}
\label{tab:obs}
\end{table*}
%%%%%%%%%%%%%%%%%%%%%%%%%%%%END OF TAB 1%%%%%%%%%%%%%%%
%

\section{The very steep spectrum radio halo in A\,697}

Abell 697 is a rich and massive cluster at z=0.282. 
It is hot ($k$T $\simeq$ 10 keV) and luminous (L$_{X}\simeq10^{45}$ erg s$^{-1}$) in the X-ray band, 
and is part of the ROSAT Brightest Cluster Sample (BCS; \citealt{ebeling98}). \\
Observational evidence shows that A\,697 is far from dynamical equilibrium: 
substructures in the galaxy distribution and in the gas have been detected through 
optical and X-ray analysis (\citealt{girardi06}; MVB10). 
Also, the absence of a cool core in the cluster was reported \citep{bauer05}.  
These studies suggest that  A\,697 is in a complex dynamical state, and it is 
probably  undergoing multiple merger/accretion of small clumps \citep{girardi06}.
\\
Diffuse radio emission at the centre of A697 was first 
suggested inspecting the NVSS and WENSS \citep{kempner01}. 
GMRT observations at 610 MHz, as part of the GMRT radio halo survey, 
allowed to confirm the presence of a giant radio halo. \\
In MBV10%\citealt{maca10}
, we studied the spectral properties of the radio halo, 
by using deep 325 MHz GMRT observations, together with the 610 MHz GMRT data and 
VLA archival observations at 1.4 GHz. 
Our multifrequency analysis showed that the integrated radio spectrum of the halo is very steep, 
with $\alpha^{1.4 \rm{GHz}}_{325 \rm{MHz}} \approx 1.7-1.8$. 

\section{Data reduction and preliminary results at 150 MHz}
\label{sec:prel}

The 150 MHz dataset of A\,697 was reduced and analysed using the 
NRAO Astronomical Image Package (AIPS). 
Sources  3C 147 and 3C 286 were observed at the 
beginning and the the end of the observing run, respectively, and were used as  flux density and bandpass calibrators. 
The source 0735+331 was used as phase calibrator. 
We applied a standard calibration procedure to the data. 
Data were affected by strong radio frequency interference (RFI), which was removed 
semi-automatically. 
\\
In each stage of self-calibration, we used the wide field imaging technique, 
to minimize the effect of non coplanar baselines. 
We covered a field of view of $\sim$5$\times$5 square degrees, with 121 facets (each wide $\sim$0.6$^{\circ}$). 
This allowed to remove the sidelobes of strong sources  far away from the field centre, 
outside the GMRT primary beam ($\sim$3$^{\circ}$). 
A few rounds of phase-only self-caibration were performed, and further editing of 
bad data was needed. \\
Preliminary clean images were then produced. 
The 1$\sigma$ rms level reached in the full resolution image is in the range 0.8-1 mJy b$^{-1}$. 
However, the presence of strong point sources in the field limits the dynamic range around them. 
The uncertainty in the calibration of the absolute flux density scale is $\sim$ 15\%. 
%
%
%%%%%%%%%%%%%%%%%%%%%%%%%%%%FIG 1%%%%%%%%%%%%%%%
\begin{figure}[htbp!]
\resizebox{\hsize}{!}{\includegraphics[clip=true]{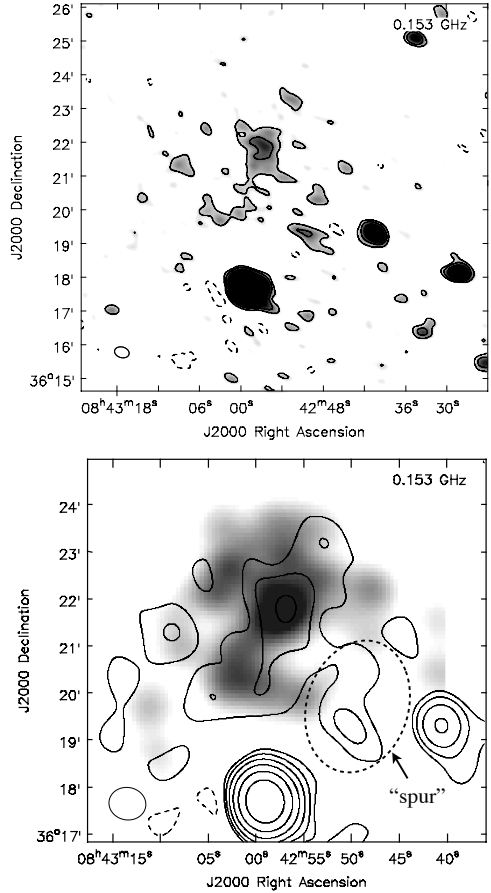}}
\caption{\footnotesize
\textit{Top panel} GMRT 150 MHz image of the central field of A\,697. 
The resolution is $\sim$ 26$^{\prime\prime}\times19^{\prime\prime}$. Contours are spaced by a factor 2, starting from $\pm$3 $\sigma$ level (i.e. $\sim$ 3 mJy b$^{-1}$). 
\textit{Bottom panel} GMRT 150 MHz contours of the radio halo in A697, overlaid on the GMRT 325 MHz image (grey scale);  the two images has the same restoring beam ($\sim$ 47$^{\prime\prime}\times41^{\prime\prime}$). 
Contours starts from $\pm$2.5 $\sigma$ level (i.e. $\sim$ 5 mJy b$^{-1}$), and are spaced by a factor 2.
} 
\label{fig:fig1}
\end{figure}
%%%%%%%%%%%%%%%%%%%%%%%%%%%%END OF FIG 1%%%%%%%%%%%%%%%
%
%
\\
\indent
The top panel of Fig. \ref{fig:fig1} shows the preliminary full resolution 
($\sim$ 26$^{\prime\prime}\times19^{\prime\prime}$) contours of the central 
12$^{\prime}\times$12$^{\prime}$ region centered on A\,697 ($\sim 3 \times 3$ Mpc$^2$, i.e. about the half cluster virial radius; see MBV10).%\citealt{maca10}).  
The rms level in the image  is 1 mJy b$^{-1}$. 
The diffuse radio emission associated with the radio halo is clearly visible around the cluster centre. 
\\
In the bottom panel of Fig. \ref{fig:fig1}, a low resolution image of the radio halo is  
shown as contours (starting from $\pm$2.5$\sigma=5$ mJy b$^{-1}$). This is overlaid to 
the 325 MHz GMRT image (in grey scale), obtained after the subtraction of discrete radio sources (same as in Fig. 2 of MBV10). %\citealt{maca10}). 
The two images has the same angular resolution ($\sim$ 47$^{\prime\prime}\times41^{\prime\prime}$). 
The radio halo at 150 MHz is mainly elongated in the South-East/North-West direction. 
Moreover, positive residuals of emission indicate an extension in the East-West direction.  
It is very extended, with largest linear size of $\sim$1.3-1.4 Mpc. 
Compared to the 325 MHz image, the halo has a similar morphology in the central $\sim1$ arcmin, and also shows a similar feature in the southern part. 
However, an accurate comparison requires subtraction of the discrete sources 
at the cluster centre. This procedure is in progress. 

\begin{figure}[htbp!]
\resizebox{\hsize}{!}{\includegraphics[clip=true]{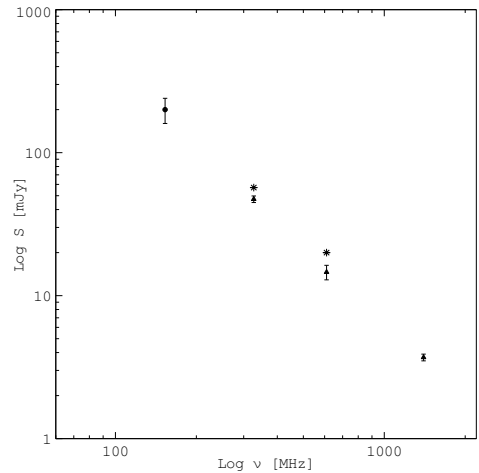}}
\caption{\footnotesize
Integrated radio spectrum of the halo. The flux density values at 325, 610 and 1400 MHz (triangles and stars) 
are from MBV10; %\citealt{maca10}; 
the circle is the preliminary value at 150 MHz. 
The errorbar represents the upper and lower values given in the text (see Sect. \ref{sec:prel}).    
} 
\label{fig:fig2}
\end{figure}

Although the image is preliminary, and further work is necessary to improve its quality, 
an estimate of the flux density of the halo at 150 MHz can be given. 
The contribution of discrete sources embedded into the diffuse halo emission (sources A, D, G in MBV10) 
%\citealt{maca10}) 
is estimated to be $\sim 20$ mJy at this frequency.  
This value was subtracted from the total flux density measured  
by integrating the low resolution image over the halo region. 
Thus we obtain S$_{150 \rm{MHz}}\sim$ 165-240, depending on whether or not we include 
the contribution of the south-western \textquotedblleft{spur}\textquotedblright\, (see Fig. \ref{fig:fig1}). 
A mean value between these two is reported in Fig. \ref{fig:fig2}, where the integrated spectrum of the halo is shown 
(same as in MBV10). %\citealt{maca10}). 
It is consistent with a steep spectral index $\alpha\sim 1.7$. 
 %
%
%%%%%%%%%%%%%%%%%%%%%%%%%%%%FIG 2%%%%%%%%%%%%%%%
%%%%%%%%%%%%%%%%%%%%%%%%%%%%END OF FIG 2%%%%%%%%%%%%%%%
%
 
\begin{acknowledgements}
GM thanks N. G. Kantharia for useful suggestions in the data reduction. 
We thank GMRT staff for their help during the observations. 
GMRT is run by the National Centre for Radio Astrophysics of the Tata Institute
of Fundamental Research. This work is partially supported by
INAF under grants PRINÐINAF2007 and PRINÐINAF2008 and 
by ASIÐINAF under grant I/088/06/0.
\end{acknowledgements}

\bibliographystyle{aa}

\end{document}